%% file: main.tex
\documentclass[10pt,journal,compsoc]{IEEEtran}

\usepackage{amsmath}
\usepackage{graphicx}
\usepackage[nocompress]{cite}
\usepackage{url}
\usepackage{hyperref}

\hypersetup{
  hidelinks,
  pdftitle={The Effect of Photorealism Consistency between the Virtual Hands and Environment on the Sense of Body Ownership and Presence in Virtual Reality},
  pdfauthor={Nami Ogawa and Takuji Narumi}
}

\newcommand{\fig}[3]{%
  \begin{figure}[tb]
    \centering
    \includegraphics[width=#1\columnwidth]{figures/#2.png}
    \caption{#3}
    \label{fig:#2}
  \end{figure}%
}

\newcommand{\figfloat}[2]{%
  \begin{figure*}[ht]
    \centering
    \includegraphics[width=\textwidth]{figures/#1.png}
    \caption{#2}
    \label{fig:#1}
  \end{figure*}%
}

\newcommand{\anovaETAbody}[6]{{$F(#1,#2)\,=\,#3$, $p\,#4\,#5$, $\eta_{p}^{2}\,=\,#6$}}

\title{The Effect of Photorealism Consistency between the Virtual Hands and Environment on the Sense of Body Ownership and Presence in Virtual Reality}

\author{%
  Nami Ogawa\thanks{This work was conducted while the first author was affiliated with DMM.com.}%
  ~and~Takuji Narumi%
  \IEEEcompsocitemizethanks{%
    \IEEEcompsocthanksitem Nami Ogawa: DMM VR lab / The University of Tokyo\protect\\
    \IEEEcompsocthanksitem Takuji Narumi: The University of Tokyo%
  }%
}

\newcommand{\ieeeacceptednotice}{%
  \parbox[t]{0.90\textwidth}{\centering\fontsize{6}{7}\selectfont
    This article has been accepted for publication in IEEE Transactions on Visualization and Computer Graphics. This is the author's version which has not been fully edited and\\[-0.2ex]
    content may change prior to final publication. Citation information: DOI 10.1109/TVCG.2026.3686249}%
}
\IEEEpubid{%
  \parbox{\textwidth}{\centering\tiny
    \textcopyright\ 2026 IEEE. All rights reserved, including rights for text and data mining and training of artificial intelligence and similar technologies. Personal use is permitted, but republication/redistribution requires IEEE permission. See \url{https://www.ieee.org/publications/rights/index.html} for more information.}%
}

\begin{document}

\IEEEtitleabstractindextext{%
  \begin{abstract}

\input{00_abst.tex}

  \end{abstract}
}

\maketitle
\IEEEdisplaynontitleabstractindextext

\input{01_body_revised.tex}

\bibliographystyle{IEEEtran}
\bibliography{references}

\begin{IEEEbiography}[{\includegraphics[width=1in,height=1.25in,clip,keepaspectratio]{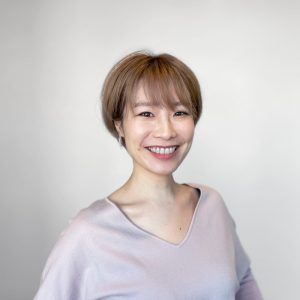}}]{Nami Ogawa}
received the B.A. degree in experimental psychology in 2015, the M.A.Sc. degree in interdisciplinary information studies in 2017, and the Ph.D. degree in engineering in 2020, all from the University of Tokyo, Tokyo, Japan.

From 2020 to 2022, she was a Principal Research Scientist with DMM VR lab Research, DMM.com, where she led research projects on self-avatar representation in virtual environments, and concurrently a Joint Researcher with the Virtual Reality Educational Research Center, the University of Tokyo. Since 2022, she has been a Research Scientist with CyberAgent AI Lab, Tokyo, Japan. Her research interests include human--computer interaction and virtual reality, with a particular emphasis on the intersection of VR and experimental psychology. Her recent work also explores the understanding and augmentation of human creativity with AI.

Dr. Ogawa was a recipient of the 2022 IEEE VGTC Virtual Reality Best Dissertation Award. She is a Member of the IEEE and the ACM.
\end{IEEEbiography}

\vspace{-2.0cm}
\begin{IEEEbiography}[{\includegraphics[width=1in,height=1.25in,clip,keepaspectratio]{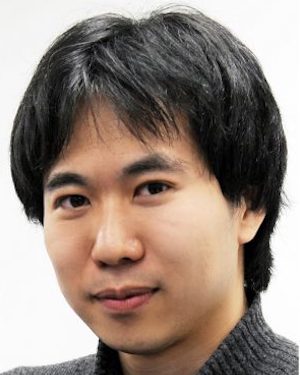}}]{Takuji Narumi}
is a professor at the Graduate School of Information Science and Technology, the University of Tokyo. His research interests include perceptual modification and human augmentation with virtual and augmented reality technologies. More recently, he is directing the Ghost Engineering Project that aims at utilizing the effect of virtual body on our mind and cognitive functions to encourage better communication between people.
\end{IEEEbiography}

\end{document}

%% file: 00_abst.tex
Virtual reality (VR) technology allows users to feel a virtual body as if it was their own (i.e., body ownership illusion).
Previous studies have explored how the visual realism of the user's virtual body influences the sense of body ownership in VR, focusing on the aspect of anthropomorphism.
However, the effect of photorealism, another element that characterizes the visual realism of a virtual human, has not been systematically examined in the context of body ownership illusion.
Therefore, we investigate the effect of the rendering style of virtual hands on the sense of body ownership, hypothesizing that the effect is affected by the rendering style of the virtual environment.
In addition, we examine the effect of photorealism on presence (i.e., the sense of being there), as existing studies have offered inconsistent evidence.
To this aim, we conducted a 3 $\times$ 3 mixed-design remote VR experiment (N=117) that factored in the rendering styles of both the virtual hand and the environment and analyzed the subjective data from the questionnaire.
The results suggest that neither the rendering style of virtual hands nor the consistency of the rendering style between the virtual hand and the environment influenced the sense of body ownership. Nevertheless, the more photorealistically the virtual environment was rendered, the stronger the presence.
Our results indicate that different dimensions of virtual body realism affect the sense of body ownership differently, thereby highlighting the importance of applying a suitable classification of realism to understand the body ownership illusion for virtual bodies more in depth.

%% file: 01_body_revised.tex
\IEEEraisesectionheading{\section{Introduction}\label{sec:introduction}}
\IEEEPARstart{A}{n}
avatar, or a virtual body controlled by users, is a key aspect of virtual reality (VR) because it is the self-representation of the user in a virtual environment.
In avatar research, there has been a growing interest in the body ownership illusion (BOI), in which users experience the illusion that the virtual body is their own body. This sensation is referred to as the sense of body ownership.
Specifying the factors influencing the BOI is important in VR because avatar embodiment not only improves the general user experience~\cite{Slater1994-ld}, but can also be used to alter the user's social cognition and behavior for good through embodied perspective-taking, one of the most powerful applications of VR~\cite{Rueda2020-ok}.

Evidence from previous studies (e.g.,~\cite{Argelaguet2016-si, Lin2016-sb, Schwind2018-uu, Yuan2010-ep,Ogawa2019-bw}) has suggested that the visual realism of a virtual body affects the sense of body ownership using the virtual hand illusion (VHI), which is the BOI toward a virtual hand that moves in spatio-temporal synchronization with one's real hand~\cite{Sanchez-Vives2010-cu}.
A considerable amount of the empirical literature outside VR has also recognized the critical role of 
the appearance of a fake body in BOIs, such as shape and texture~\cite{Tsakiris2010-uq, Haans2008-an,Kilteni2015-jd}.
That is, if a virtual hand is semantically congruent in terms of its resemblance to a human hand, it is likely to elicit the VHI~\cite{Kilteni2015-jd}.

By contrast, researchers in different fields have studied human reactions and perceptions toward virtual humans with different visual realism in the pursuit of making lifelike virtual humans~\cite{Nowak2003-ag,McDonnell2012-oo, Zibrek2014-hc, Zell2015-bx,  Zibrek2018-ql, Zibrek2019-xm, Schindler2017-sm}.
Here, virtual humans refer to human models using three-dimensional computer graphics (3DCG) controlled by humans (i.e., avatars) or by computers (i.e., agents)~\cite{Nowak2003-ag,Garau2003-tf}.
In this line of research, visual realism, sometimes called visual fidelity, is considered to consist of multiple dimensions.
Garau~\cite{Garau2003-tf} classified the visual fidelity of avatars into three dimensions: anthropomorphism (i.e., morphological human-likeness), photorealism (i.e., level of visual detail), and truthfulness (i.e., resemblance between the user and avatar).

However, the elements that constitute realism remain unclear and undifferentiated in previous VHI research, even though the VHI embodies a virtual human hand model.
Specifically, in previous studies examining the influence of realism on the VHI, realism has mainly referred to the anthropomorphism dimension alone (e.g., \cite{Argelaguet2016-si, Yuan2010-ep,Ogawa2019-bw}).
Only recently have several studies started to investigate the effect of the truthfulness dimension and found that the personalization of a virtual hand or full body increases body ownership more than a generic one~\cite{Jung2018-wx,Waltemate2018-cf,Gorisse2019-zk}.
However, although several previous studies have manipulated both the anthropomorphism and the photorealism of virtual hands~\cite{Schwind2017-hi,Lin2016-sb}, 
the findings on how photorealism affects the VHI remain inconsistent.
This background motivated us to investigate the effect of the photorealism of virtual hands on the VHI.
In this paper, we manipulated photorealism by varying the rendering style, specifically materials and textures, from realistic to cartoon-like along an axis of indistinguishability from a photograph. This manipulation allowed us to isolate the effects of photorealism while holding other components of realism constant, namely anthropomorphism (morphological shape) and truthfulness (personalized resemblance).

To identify the prospective moderator variable and hypothesize the effect of the photorealism, we
highlight studies of the effect of photorealism on users' perceptions in virtual environments.
Research recognizes the important role played by the photorealism of virtual environments on presence (i.e., the sense of being there), although the evidence is inconsistent~\cite{Zimmons2003-cj,Mania2004-mm,Vinayagamoorthy2004-lk,Khanna2006-nu,Slater2009-ix, Yu2012-kl}. 
Zibrek et al.~\cite{Zibrek2019-xm} indicated users' potential adverse reaction to the inconsistency of the rendering style between a virtual human and the environment.
Furthermore, in AR systems, non-photorealistic renderings are used to achieve the perceptual unification of a scene that includes virtual and real content~\cite{Steptoe2014-ge,Haller2004-he,Fischer2006-hp}.
Together, these studies provide important insights into the overall consistency of photorealism.

\begin{figure*}[t]
\begin{centering}
  \includegraphics[width=2\columnwidth]{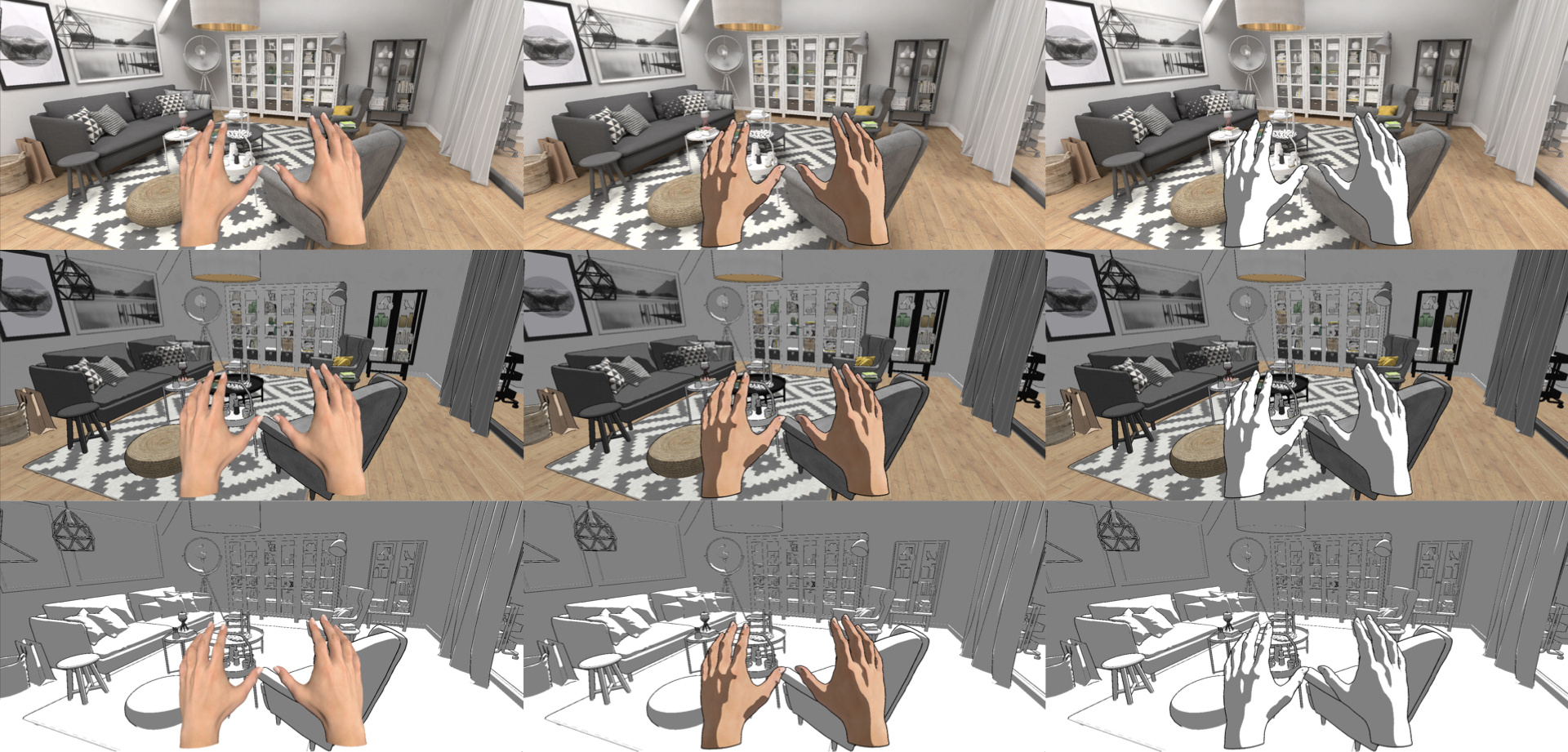}
  \caption{The combination of virtual hands and environments with different rendering styles used in this study. From left (top) to right (bottom): realistic, toon, and sketch.
  }
  \label{fig:teaser}
\end{centering}
\end{figure*}

Therefore, we aim to answer the following research question:
\textit{Does the photorealism of a virtual hand affect the VHI? In particular, does the consistency of photorealism between the body and environment affect users' perceptions such as the sense of body ownership and presence?}
Based on existing research, we propose the following hypotheses.
\textbf{[H1, H2]} The consistency of the photorealism between the body and environment increases the sense of body ownership and presence.
\textbf{[H3, H4]} The more photorealistic the virtual hands (environment), the stronger the sense of body ownership (presence).
To test these hypotheses, we conducted a 3 $\times$ 3 mixed-design remote VR experiment (N=117; sampled from consumer HMD owners) that factors in the rendering styles of both the virtual hand and the environment (\autoref{fig:teaser}).
The results suggest that neither the photorealism of virtual hands nor the consistency of photorealism influences the VHI, rejecting the hypotheses. Nevertheless, the more photorealistically the virtual environment was rendered, the stronger the presence.
Together, our study contributes to the literature in the following ways.

\noindent \quad$\bullet$
We systematically examine the effects of photorealism alone as a dimension of realism, explicitly dissociating it from other contributing factors such as anthropomorphism and truthfulness. This extends previous VHI studies by showing that different dimensions of virtual body realism may have different impacts on the sense of body ownership.

\noindent \quad$\bullet$ 
By examining the interaction of photorealism between the body and environment, we add a new perspective to the study of BOIs, namely, how environmental information
affects the sense of body ownership.

\noindent \quad$\bullet$ 
Our relatively large controlled experiment provides additional empirical findings into the independent effects of body and environmental photorealism on the sense of body ownership and presence, respectively, for which the evidence has thus far been inconsistent.

\section{Related Work}

\subsection{Photorealism: A Component of Realism}
To situate photorealism within the broader literature, we first outline how realism has been conceptualized and classified in VR and computer graphics. In immersive systems, realism broadly refers to the extent to which a virtual experience meets our expectations about the real world; although terminology varies across authors, usage converges on closely related meanings \cite{Goncalves2021-av,Goncalves2023-ne,Goncalves2025-sf}. A common distinction separates subjective (phenomenological) and objective (stimulus driven) realism. Photorealism falls under objective realism. Objective realism is often treated as essentially equivalent to fidelity \cite{Goncalves2023-ne}.

As with definitions, the dimensions and components of realism also vary across the literature (for a systematic review, see \cite{Goncalves2023-ne}). Nonetheless, many accounts agree that objective realism includes rendering aspects such as surface and material depiction and shading and illumination. For example, in Perroud et al.’s classification of five types of realism, “Realistic looking” refers to detailed shaders, materials, and lighting. In computer graphics, Ferwerda \cite{Ferwerda2003-kw} distinguishes photorealism, defined as indistinguishability from a photograph, from other notions of realism, namely physical realism and functional realism.

Although there is no common taxonomy for objective realism, we specifically draw on Garau’s \cite{Garau2003-tf} classification of the visual fidelity of avatars, which comprises three dimensions: anthropomorphism (non humanoid versus humanoid), photorealism (cartoonish versus visually detailed), and truthfulness (does not resemble versus resembles the user). This classification originates from the context of social interactions with an avatar in VR and explicitly considers components relevant to virtual humans. Although its direct application to the VHI should be considered with care, as discussed later, anthropomorphism is implicitly manipulated as a means of operationalizing realism in most VHI research. We therefore believe that clearly separating these dimensions will advance the systematization of research in this field.

In this paper, we interpret this classification for virtual hands as follows: photorealism refers to texture and material fidelity and to shading and illumination realism; anthropomorphism refers to shape and morphology; and truthfulness refers to user resemblance (personalization). Achieving high photorealism in real time often requires substantial computational resources, so increasing photorealism is not always the best choice in VR, where frame rate and latency are critical \cite{Goncalves2023-ne,Haller2004-he}. This motivates examining when and how photorealism benefits user experience, given the practical tradeoffs developers often face.

\subsection{The VHI}

\subsubsection{Top-down Factors in the VHI}
The BOI refers to the phenomenon that humans feel body ownership over a non-corporeal object under certain conditions. The VHI is a type of BOI in which a virtual hand moves in conjunction with the user's hand movements~\cite{Sanchez-Vives2010-cu,Yuan2010-ep}. While the rubber hand illusion, the classic example of the BOI, is triggered by synchronous visuo-tactile stimulation~\cite{Botvinick1998-kt}, the VHI is triggered by visuomotor synchronization. However, both are considered to be fundamentally the same phenomenon because they share many common factors, such as constraints on the appearance of a rubber or virtual hand and spatiotemporal visuo-tactile or visuomotor synchrony~\cite{Kilteni2015-jd}.

Experimental findings on the BOI, both within and outside VR, have shown that the sense of body ownership emerges from a combination of bottom-up and top-down information~\cite{Kilteni2012-ji}. Kilteni et al.~\cite{Kilteni2012-ji} summarized bottom-up information as afferent sensory input that arrives at the brain from the sensory organs, whereas top-down information consists of cognitive processes that may modulate the processing of sensory stimuli. The existence of a top-down process indicates that the BOI is sensitive to the semantic resemblance between a non-corporeal object (e.g., a rubber or virtual hand) and the human body (for a review, see~\cite{Kilteni2015-jd}).

These accounts suggest that top-down factors tied to visual realism are likely to modulate the VHI, motivating a focused examination of appearance-related dimensions. Thus, a number of VHI studies have identified and evaluated the effect of the visual realism of virtual hands. Existing research has shown that a realistic virtual human hand elicits a stronger sense of body ownership than non-human hands (e.g., iconic, zombie, or cartoon)~\cite{Argelaguet2016-si,Schwind2018-uu,Lin2016-sb} and non-corporeal objects (e.g., spheres, blocks, boards, or arrows)~\cite{Argelaguet2016-si,Lin2016-sb,Ogawa2019-bw,Yuan2010-ep}.

However, most such studies deal only with the general concept of realism. Although the realism of a virtual hand is a mixture of multiple elements, no common taxonomy for hand appearance has thus far been used in VHI studies. Hence, different studies use different hand appearances without adopting consistent terms, making it difficult to integrate findings into a unified body of knowledge. Moreover, because taxonomies of top-down factors for the general BOI are largely based on rubber hand illusion studies (e.g., shape, texture, and anatomical plausibility~\cite{Kilteni2015-jd}), they do not necessarily explain variations in virtual hand appearance. Since it remains unknown whether each dimension of realism exerts the same influence on the VHI, clearer classification and systematic research are needed.

\subsubsection{Effects of Hand Photorealism on the VHI}
Drawing on Garau's classification, realism in previous VHI studies has mainly concerned the anthropomorphism dimension (e.g., \cite{Argelaguet2016-si}).	
The truthfulness dimension has also been explored in terms of avatar personalization.	
Jung et al.~\cite{Jung2018-wx} showed that a personalized virtual hand (i.e., a video-see-through hand) increases body ownership to a greater extent than a generic 3DCG hand. Similar findings have been reported for BOIs with a full-body self-avatar from the first-person perspective~\cite{Waltemate2018-cf} as well as from the third-person perspective~\cite{Gorisse2019-zk}.	
By contrast, the photorealism dimension has not been investigated in a systematic manner in the context of the VHI, which motivates our focus on this axis.

One of the few studies to incorporate aspects of photorealism into the VHI is the work of Lin et al.~\cite{Lin2016-sb}. They investigated the VHI with six virtual hand appearances that differed in anthropomorphism (their ``realism''), rendering style (photorealism), and sensitivity to pain. However, differences in VHI strength across rendering styles (realistic, toony, very toony) varied by experimental design (within- vs.\ between-subjects) and by the questions asked. Several other studies have also used cartoon virtual hands among various appearances, but their focus has been on perceived presence, likability, and eeriness~\cite{Schwind2017-hi} as well as accuracy in a pointing task~\cite{Schwind2018-lq}, rather than directly on body ownership or the VHI. 

Taken together, the impact of photorealism on the VHI has been underexplored, and the available evidence is inconsistent.
We therefore focus on the photorealism dimension while holding anthropomorphism and truthfulness constant, enabling an examination of its specific contribution to the VHI.
We assume that the photorealism of a virtual hand alone can affect the VHI insofar as perceived realism changes, given that top-down processes modulate the BOI.

\subsection{Photorealism in VR and AR}
\subsubsection{Photorealism of Avatars}
Although little attention has been paid to photorealism in VHI research, a great deal of previous research on virtual humans using 3DCG, both within and outside VR, has explored the effect of photorealism. In such fields, people's perceptions of virtual humans with different rendering styles have been studied to understand how to render them as realistic, attractive, or expressive (e.g., \cite{Zibrek2018-ql,McDonnell2012-oo,Zell2015-bx}). For instance, McDonnell et al.\ \cite{McDonnell2012-oo} investigated the effect of photorealism on perceived realism and other feelings (i.e., appeal, familiarity, and friendliness) in facial animation of virtual humans in video.

A number of studies have examined virtual human photorealism in images or videos (e.g., \cite{MacDorman2009-hz,Schindler2017-sm,McDonnell2012-oo,Zell2015-bx}), often in relation to the uncanny valley theory, which states that subtle realism imperfections cause negative responses from perceivers \cite{Mori1970-vq,Seyama2007-oo}. Indeed, the uncanny valley effect has been confirmed, although the relationship does not seem to be as simple as originally proposed \cite{MacDorman2009-hz,Schindler2017-sm,McDonnell2012-oo}. In addition, interaction effects between the categories of realism (i.e., visual and behavioral) and between components of visual realism have been identified \cite{McDonnell2012-oo,Zell2015-bx}. For example, Zell et al.\ \cite{Zell2015-bx} showed that inconsistency between the stylization level of anthropomorphism (i.e., shape) and photorealism (i.e., material style) impacts the appeal and attractiveness of virtual humans, making them look eerie.

VR technology has raised another concern in virtual human studies: whether virtual humans are treated as if they are human in addition to being perceived as human. Such interest has led to an exploration of the impact of rendering style on emotional reactions \cite{Volonte2016-uy,Zibrek2018-ql}, visual attention \cite{Volonte2019-tn}, and proximity \cite{Zibrek2017-yb,Zibrek2018-ql,Zibrek2019-xm,Zibrek2019-ha}. In particular, the impact of virtual human realism (e.g., anthropomorphism \cite{Garau2003-ni,Mansour2006-cv,Nowak2003-ag,Bailenson2006-qb,Latoschik2017-tv} and photorealism \cite{Zibrek2018-ql,Zibrek2019-ha,Zibrek2019-xm}) on social presence and co-presence, measured either subjectively (i.e., questionnaires) or objectively (i.e., proximity), has been an important perspective in VR studies. The series of studies by Zibrek et al.\ \cite{Zibrek2017-yb,Zibrek2018-ql,Zibrek2019-xm,Zibrek2019-ha} investigated the effect of photorealism on social presence: earlier studies reported that the rendering style of virtual humans does not affect social presence levels \cite{Zibrek2017-yb,Zibrek2018-ql,Zibrek2019-xm}, whereas a more recent study that controlled the rendering styles of a virtual human, the environment, and the virtual self-body simultaneously showed effects of rendering style \cite{Zibrek2019-ha}. In addition, some of their studies measured place illusion (i.e., the sense of being there, often called presence \cite{Slater2009-hb}), finding that realistic rendering of a virtual human and the environment can increase presence \cite{Zibrek2019-xm,Zibrek2019-ha}. Taken together, photorealism has been recognized as an important element that characterizes how humans perceive virtual humans, with both absolute fidelity and cross-component consistency playing roles.

By contrast, Zibrek et al.~\cite{Zibrek2019-ha} incorporated the sense of body ownership into the context of virtual human studies. They investigated the effect of the rendering style of the entire virtual world, including a self-avatar in a full-body suit with no facial features, on the sense of body ownership as well as presence. The results suggested that rendering style does not influence the sense of body ownership.
We expand this work by systematically separating the influence of the virtual environment from that of the body (i.e., virtual hands), with a focus on the effect of their combination.
\subsubsection{Photorealism of Virtual Environments}
Beyond virtual humans, the photorealism of virtual environments also affects user experience, especially presence. Given that ``realism'' includes multiple dimensions, it is unsurprising that evidence about whether realism in general increases presence is inconclusive \cite{Jung2021-ml}. A review indicates that, overall, higher realism has a positive impact on user experience, but the effect depends on which factors are manipulated \cite{Goncalves2023-ne}.

Nonetheless, photorealism has traditionally been considered a representative dimension of environmental realism, particularly in visual terms \cite{Ferwerda2003-kw}. For example, visual realism (e.g., wireframes vs.\ shaded 3D renderings) has been considered important in the well-known virtuality continuum \cite{Milgram1994-ls}, although recent studies emphasize the importance of multisensory input and a broader concept of coherence \cite{Skarbez2021-ci,Jung2021-ml}. This contrasts with the fact that anthropomorphism has often been treated as the main factor of avatar realism.

Generally, as visual realism increases, there is a tendency for presence to increase \cite{Goncalves2021-ny,Goncalves2023-ne}. However, this is not always the case, and improvements in photorealism do not necessarily lead to enhanced presence \cite{Jung2021-ml}.
In fact, evidence of the effect of photorealism on presence remains inconsistent. Earlier studies often concluded that photorealistic rendering does not necessarily increase presence \cite{Zimmons2003-cj,Mania2004-mm,Vinayagamoorthy2004-lk}, whereas others found that some rendering elements (e.g., real-time shadows and reflections) increase presence \cite{Khanna2006-nu,Slater2009-ix,Yu2012-kl}. It has also been noted that increasing photorealism can sacrifice real-time, high-frame-rate rendering, especially when computational resources are limited \cite{Garau2003-tf,Zibrek2019-xm}. In fact, possibly owing to advances in computational power, recent studies tend to show significant positive effects of photorealism on presence \cite{Schwind2019-th,Zibrek2019-xm,Zibrek2019-ha,Hvass2017-wy}. Hence, based on recent evidence, we hypothesize that a more photorealistic environment increases presence.

\subsubsection{Effects of Photorealism Consistency}
The importance of visual consistency between scene elements has been recognized in several contexts, especially in AR. 
As an alternative approach to enhancing the photorealism of virtual objects, non-photorealistic rendering has been used in AR to reduce the photorealism of real-world view to match virtual content, thereby making them less distinguishable \cite{Haller2004-he,Steptoe2014-ge,Fischer2006-hp}.
Steptoe et al.\ \cite{Steptoe2014-ge} found no significant effect of such homogenization via non-photorealistic rendering on presence in immersive AR, yet their work highlights that cross-source consistency in photorealism can shape user perception. Although findings from AR do not directly generalize to VR, these results motivate examining the role of photorealism consistency between sources in VR.

In VR, previous studies assessing the effect of the photorealism of the virtual self-body either controlled only the body's rendering style~\cite{Lin2016-sb,Schwind2017-hi,Schwind2018-lq} or applied the same rendering style to both the body and the environment~\cite{Zibrek2019-ha}. We therefore manipulate them independently to examine a potential interaction. Specifically, we test whether a match in photorealism between the environment and the body in VR influences perception (i.e., the sense of body ownership and presence). Regarding social interactions in VR, Latoschik et al. \cite{Latoschik2017-tv} indicated that the appearance of others' avatars affects users' self-perception of their own body. This finding suggests that not only the appearance of a virtual body itself but also contextual information can influence self-perception in a top-down manner. Therefore, we hypothesize that consistency in rendering style between the body and the environment can increase the sense of body ownership and presence by homogenizing the overall appearance of the scene.

To the best of our knowledge, no study has thus far investigated the moderation (interaction) effect of the environment on BOI factors. This is understandable, given that much BOI research has been conducted in real environments, which are not usually subject to control. However, when designing VR experiences, whether for practical use or research purposes, it is necessary to determine not only the design of the body but also the design of the environment at the same time. From a practical standpoint, our study aims to provide insights into common concerns in the design of VR, including: \textit{``Can we decide the rendering style of the body and environment independently?''} and \textit{``Is one rendering style of the body particularly (in)sensitive to the rendering style of the environment with respect to body ownership?''} At the same time, from the perspective of BOI research, our study adds a perspective on how environmental information affects the sense of body ownership. 

\section{Experiment}
To investigate the aforementioned research questions, we systematically controlled the level of photorealism of virtual instances by manipulating rendering styles, explicitly separating it from other components of visual realism of a virtual avatar (e.g., anthropomorphism and truthfulness). In addition, we controlled the photorealism of virtual hands and environments independently, examining both independent and combined effects.

The experiment was conducted in the wild, i.e., data were collected on consumer-owned devices. As it has been argued that out-of-lab studies can take advantage of the collection of larger-scale datasets by making efforts to improve data reliability compared to a conventional laboratory experiment \cite{Steed2016-bn}, we carefully conducted pilot testing with the help of quality assurance (QA) testers who were used to test video games. Additionally, we carefully checked the data quality and excluded suspected uncompliant data (see Section~\ref{subsec:results-data} and Appendix).

\subsection{Design and Hypotheses}

\textbf{\textit{Design.}}
Our experiment investigated the effect of the photorealism of a virtual body and the environment, especially the consistency between them, on the sense of body ownership and presence.
The study followed a 3 $\times$ 3 mixed factorial design.
The independent variables were the photorealism of a virtual hand (between-factors: \textit{realistic, toon, and sketch}) and photorealism of the environment (within-factors: \textit{realistic, toon, and sketch}).
The main dependent variable was the subjective evaluation of the sense of body ownership and presence.
Participants wearing the head mounted display (HMD) performed a simple pointing task in three virtual rooms that differed only in rendering style and evaluated how they perceived their body and the environment in each room. The experiment was conducted remotely.

\noindent
\textbf{\textit{Hypotheses.}}
Our hypotheses were as follows:

\noindent
\textbf{[H1]} When the virtual hands and environment are rendered in the same style, the sense of body ownership increases compared with when rendered in different styles.

\noindent
\textbf{[H2]} When the virtual hands and environment are rendered in the same style, the sense of presence increases compared with when rendered in different styles.

\noindent
\textbf{[H3]} The more photorealistically the virtual hands are rendered, the stronger the sense of body ownership.

\noindent
\textbf{[H4]} The more photorealistically the virtual environment is rendered, the stronger the sense of presence.

\textbf{H1} and \textbf{H2} are the main interests in this study, as no study has thus far investigated the effect of consistency of elements in VR. Based on the literature on photorealism in virtual humans and in AR, we hypothesized that the consistency between the body and environment could improve body ownership and presence in VR as well. 
As we were primarily interested in whether or not the consistency effect exists, we did not make any prior assumptions about each partial effect of the interaction.
In other words, we examined whether any effects exist where the body ownership (presence) changes merely by changing the environment (hand), despite the same appearance of the virtual hands (environment).
\textbf{H3} was formulated considering that photorealism is an element of realism and that realistic hands have been shown to strengthen the VHI (e.g.,~\cite{Argelaguet2016-si,Schwind2018-uu, Lin2016-sb,Ogawa2019-bw, Yuan2010-ep}). Nonetheless, a previous study~\cite{Lin2016-sb} only partially corroborated the effect in a manner that depended on the experimental protocol.
\textbf{H4} was formulated to be consistent with the effect shown in recent studies~\cite{Schwind2019-th, Zibrek2019-xm,Zibrek2019-ha,Hvass2017-wy}, although earlier studies have shown inconsistent results (e.g.,~\cite{Zimmons2003-cj,Mania2004-mm,Vinayagamoorthy2004-lk,Khanna2006-nu,Slater2009-ix, Yu2012-kl}).
\subsection{Apparatus and Participants}
\label{subsec:participants}

Having an Oculus Quest (either 1 or 2), having proficiency in Japanese, and being over the age of 18 were required conditions to participate.
We limited the device to only Oculus Quest because we wanted to use its hand tracking feature.
Oculus Quest 1 and 2 had $1440 \times 1600$ and $1832 \times 1920$ pixels per eye, respectively. Both had a refresh rate of 72Hz.
We developed the application using Unity version 2019.4.2f1. The application was built and distributed as an Android Application Package.
The experiment was approved by the ethical committee in the Graduate School of Information Science and Technology, The University of Tokyo.

In total,
128 individuals completed the application.
Among them, 124 answered the post-experiment demographic survey.
Altogether, 115 participants (92.7\%) identified as male, while the rest identified as female. 
Although we also provided the option of ``Not Listed,'' no participant chose it.
A total of 116 participants (93.5\%) identified as right-handed, while the rest identified as left-handed.
Eighty-five participants (69.1\%) answered that they used Oculus Quest 2, while the rest answered Oculus Quest.
Age varied from 18 to 68 (31.04 $\pm$ 9.99 SD years old).
The distributions of the participants' previous VR experience are shown in Appendix Table 1.

\subsection{Manipulation of Rendering Style}
We chose three rendering styles based on Zibrek et al.~\cite{Zibrek2019-xm}, while also referring to other studies~\cite{Lin2016-sb,Schwind2017-hi,McDonnell2012-oo,Steptoe2014-ge}.
To manipulate the levels of photorealism so that the three levels could be perceived as different as possible, we first chose the most photorealistic (i.e., detailed in mesh and material) models possible that could run smoothly (i.e., over 40 fps) in Oculus Quest.
Thus, the following 3D models were used under \textit{realistic} conditions.
For the environment, we used ArchVizPRO Vol. 6\footnote{\url{https://oneirosvr.com/archvizpro-vol-6/}.}, provided by Oneiros srl. 
For the hand, we used a model obtained from the Leap Motion Software Development Kit (SDK) version 2.3\footnote{\url{https://developer-archive.leapmotion.com/documentation/v2/unity/unity/Unity_Hand_Assets.html}.} as models from the SDK have been used in several previous studies as a realistic virtual hand (e.g., \cite{Schwind2017-hi,Argelaguet2016-si,Lin2016-sb,Ogawa2019-bw}). 
Of these, we adopted a model with little maleness based on the previous findings on gender differences in perceiving virtual hands that women dislike male virtual hands while men accept and feel presence with virtual hands of both genders~\cite{Schwind2017-hi}.
The adopted model was felt realistic enough for both men and women without being eerie or repulsive. 

To create the virtual stimuli in the medium level of photorealism, we stylized the photorealistic models by applying toon shading. These stimuli represented the \textit{toon} conditions.
We employed toon shading because it has been used to manipulate rendering styles in many previous studies~\cite{Lin2016-sb,Schwind2017-hi, Schwind2018-lq, McDonnell2012-oo}.
In addition, applying the same shaders for the body and environment, rather than manually adjusting the material parameters, ensured the visual match of the rendering styles for the corresponding conditions. 
To implement the toon shading, we added a black outline using a wire-frame method and replaced the continuous shading with two-color shading, but retained the texture information.

To create the \textit{sketch} conditions, the lowest level of photorealism, we also applied the toon shader on the photorealistic models, but removed the texture information.
Such a gray-scale environment with an outline has been used in previous studies of photorealism~\cite{Zibrek2019-xm,Steptoe2014-ge}.
Several studies have mentioned that completely removing the shading information could disturb depth perception~\cite{Zibrek2019-xm,Schwind2018-lq}. Thus, we considered an outline with two-color shading as the least photorealistic models that could be practically used in VR.

\subsection{Pointing Task}
\label{subsec: pointing}
The pointing task was designed so that the participants could naturally see their body and the environment at the same time for a certain period.
Because participants' serious engagement is not always assured when an experiment is conducted remotely without the interjection of an experimenter, we carefully designed the task, although we were not interested in task performance itself.
Indeed, if we simply instruct them to look at a virtual body for a while, they will likely be too bored to ignore the instructions, or the instructions will be so direct that they notice the hypothesis of the experiment.
On the contrary, videogame-like absorbing tasks could interfere with the effect of photorealism.
Indeed, while developing an informal prototype, we verified that VR-specific interactions that would not happen in the real world (e.g., a floating menu and break in physical law) could feel strange particularly in realistic conditions, which could considerably decrease the sense of body ownership and presence only in the realistic conditions. 
This is in line with the literature showing an interaction effect between behavioral and visual realism (i.e., fidelity) on perceiving virtual humans~\cite{Garau2003-tf,Garau2003-ni,Chaminade2007-bt}.

For those reasons, in addition to the time needed just to move and look around freely (called \textit{pre- and post-embodiment}), we adopted a pointing task in which the participants were asked to search for a target object in the virtual room and point at it continuously for three seconds.
The participants could use either hand to point at an object to ease fatigue.
The pointing gesture was detected by ray casting methods based on the direction of the index fingertip \cite{Corradini2002-zu, Schwind2018-lq}.
The target object was randomly chosen from 18 items that existed in the virtual room, and its image was displayed on the screen of a computer that also existed in the room.
When the invisible ray cast intersected with the object for three seconds, a sound was played to signal to the participant that the pointing was a success. No other feedback was given to the participants because we wanted to remove VR-specific interactions from the task scene as much as possible.
As soon as the sound played, the target image on the virtual screen was changed to the next one.
Targets were chosen so that there was no duplication until all 18 items appeared once as targets, even when the environment condition changed. If the participants had completed a total of 18 trials, the next round was repeated similarly. That is, from the 19th trial onwards, targets were again chosen in random order from the same set of 18 items.
Although we were not interested in task performance, we manually adjusted the ease of pointing to all the objects to make it as uniform as possible so that the difficulty of the task would be similar among the three environments even though the targets were chosen in a random order.
Eventually, the participants (N=117) completed 35.15 $\pm$ 12.29 (SD) items on average per experiment (i.e., 90 s $\times$ 3 conditions), which indicates that 7.81 s $\pm$ 2.86 s (SD) was spent on average for each item. In other words, the average participant found the same item approximately twice throughout the experiment.

\subsection{Measurement}
\label{measurement}
We collected the answers to the questionnaire, duration spent answering each item, task performance of the pointing task (i.e., the time and number of successful trials), and state of hand tracking (i.e., whether 
the virtual hands were displayed in sight).
We analyzed the answers to the questionnaire to test the hypotheses but did not analyze the task performance data, as noted above. We used the other data to determine whether the participants had completed the experiment properly.
All the data was anonymized and uploaded to a server.

\input{02_questionnaire_table}

The subjective evaluation of the sense of body ownership and presence was assessed through a questionnaire within the HMD (see \autoref{table:questionnaire} for all the items).
All the questionnaire items were translated into Japanese.
Each response was scored on a seven-point Likert scale from $-3$ to $+3$.
For the sense of body ownership, we used
three items from the avatar embodiment questionnaire~\cite{Gonzalez-Franco2018-ln} by omitting items that were not applicable to our study context, such as the question on tactile sensation.
For presence, we used the Slater--Usoh--Steed questionnaire~\cite{Usoh2000-yf}'s
four items out of six, following the experiment conducted in the wild by Steed et al.~\cite{Steed2016-cz}.
In addition, we measured the perceived realism of the hands and environment following previous studies~\cite{Zibrek2019-xm,Steptoe2014-ge,Lin2016-sb} to understand whether the manipulation of rendering style was sufficient to make a difference to perceived realism and whether the same manipulation of rendering style had a different effect on perceived realism between the body and environment.

Since we allowed anyone to participate in the experiment with the inducement of a monetary reward and anonymously, we made special efforts to ensure the reliability of the responses~\cite{Meade2012-pn} (see \autoref{subsec:results-data} for the details).
Thus, to measure consistency, we inverted the axis of half of the presence items (i.e., Q6 and Q7). That is, we changed the direction of the axis of these items so that $-3$ indicated the strongest sense of presence, as opposed to $+3$ indicating the strongest in the original questionnaire. The purpose of this change was to distinguish participants who were simply repeating similar responses without carefully reading the question text from those who were repeating similar responses because they actually felt the strong or weak sense of presence.
By contrast, as the sense of body ownership items originally included control items, we kept the original statement and axes.
In addition, we recorded the response time to each question, as those participants with extremely short response times were unlikely to be following the instructions correctly.

Since it was difficult to create an easy-to-use interface for the questionnaire using the hand tracking system without potentially influencing the sense of body ownership and presence (see \autoref{subsec: pointing}), the questionnaire was answered in a different simple environment from the one being evaluated (\autoref{fig:neutral}). The appearance of the virtual hands was also made neutral (i.e., semi-transparent blue texture with simplified mesh) during the questionnaire so that the participants would not confuse the ownership at that time with the ownership being evaluated (i.e., in the embodiment phase).

\fig{1}{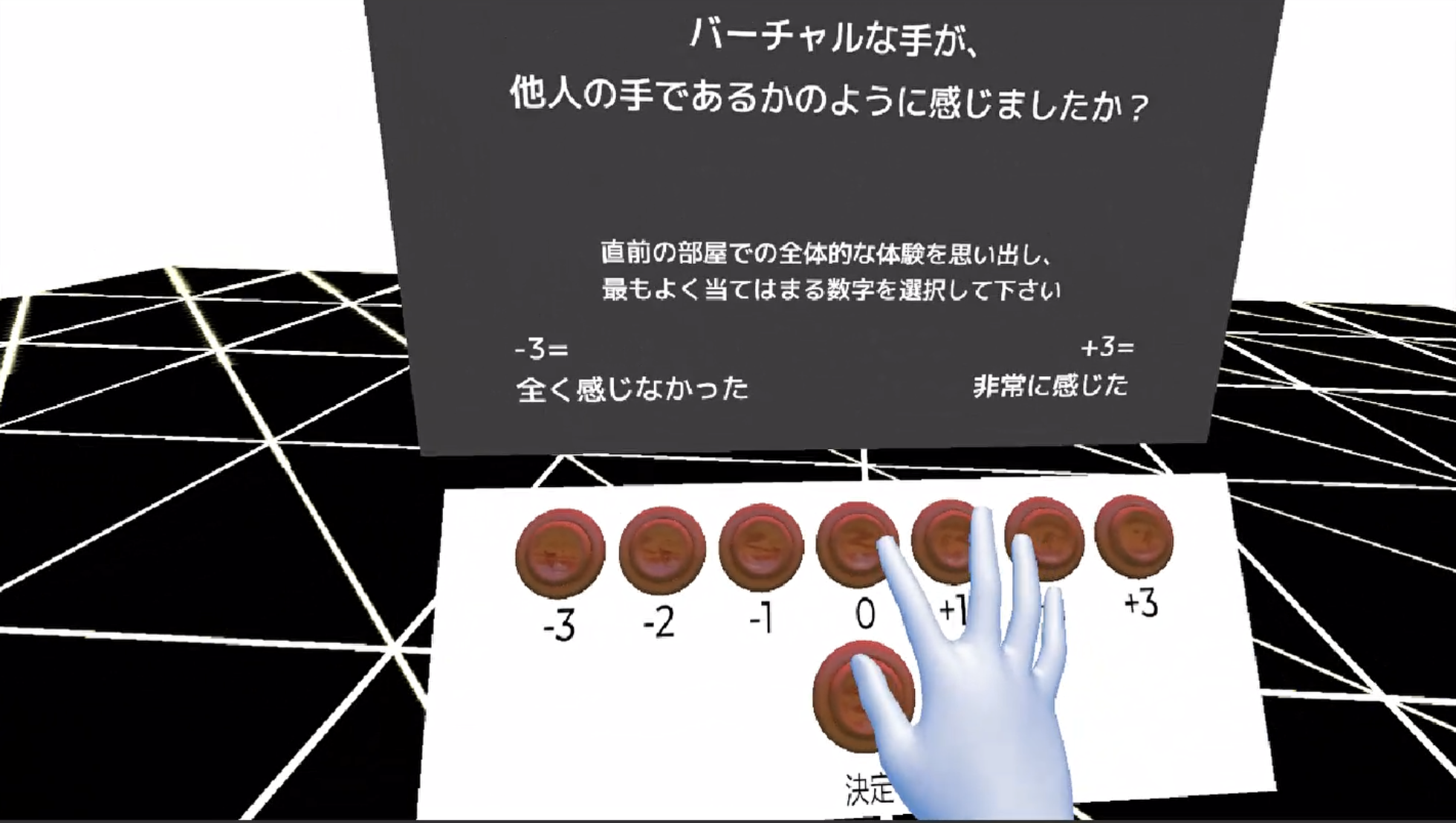}{A screenshot of the questionnaire phase. Neutral hands with a simple environment were used.
}

\subsection{Procedure}
\label{sec:flow}
\fig{1}{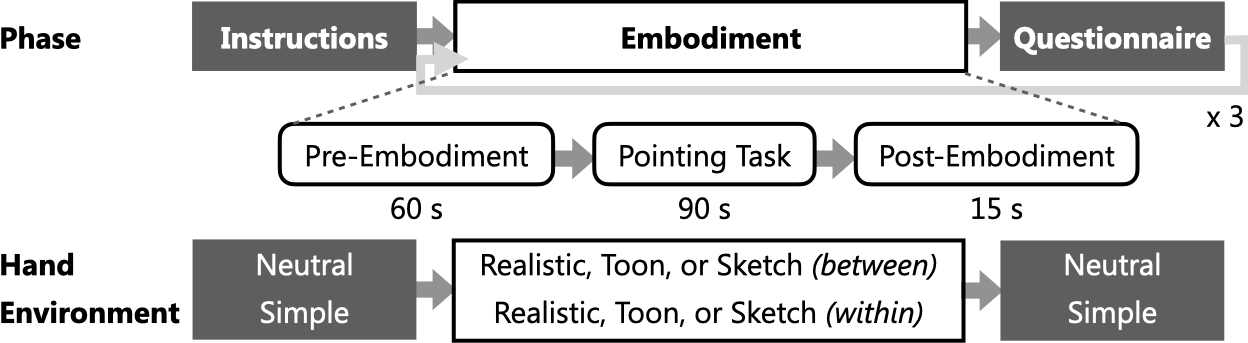}{Overview of the experiment.
}

The procedure was divided into (a) web-based instructions and setup, (b) within-VR experiment, and (c) web-based post-survey. 
An overview is illustrated in \autoref{fig:diagram}.
The whole process took approximately 0.5 to 1 hours to complete.

\subsubsection{Participant Recruitment and Pre-experiment Information}
We recruited the participants for the experiment through social media (i.e., Twitter).
Anyone interested in participating in the experiment could beforehand read the information (e.g., ethics information, requirements, rewards, and general procedures) on the website of this study~\footnote{\url{https://dmm-com.github.io/vrlab-research-web/expt/2020-1st/}}.
The website presented only a brief aim of the experiment (i.e., research on user experience when using hand tracking in VR), with no explicit mention of rendering styles.
By providing an approved statement of consent, the participants could download the application from the website.
This process took 10--30 minutes.

\subsubsection{Experience in VR Application}
The participants installed and executed the application themselves. 
For health and safety reasons, they were required to be in a sitting position while playing the application.

\textbf{Instructions Phase.}
When the participants ran the application, they appeared with neutral virtual hands in a simplified environment.
They could push the virtual buttons with the index finger of their virtual hand.
After they practiced how to push the buttons and point at an object, they read the instructions and started the main experiment. 

\textbf{Embodiment Phase.}
They were randomly assigned to each hand condition as equally as possible.
They appeared with the allocated virtual hands in a room rendered in any one of the environment photorealism settings.
In the room, all the instructions and the remaining time were displayed on a monitor on the table.
The first instruction was "Feel free to look around the space, move your body (virtual hands) around, and become familiar with the space and your body," which shaped \textit{Pre-embodiment} for 60 seconds.
Next, they were asked to conduct a pointing task, which lasted for 90 seconds.
Finally, they were given another 15 seconds with the instruction, "You will be asked questions about your experience and impressions in this room," which shaped \textit{Post-embodiment}.

\textbf{Questionnaire Phase.}
The virtual hands and environment were the same as in the instructions phase.
A question from the questionnaire and buttons with numbers from $-3$ to $+3$ were displayed.
The participants were instructed to answer each question based on the recall of the experience of the preceding scene and not based on the impression of the virtual hands or environment they were currently observing.
The order of the nine questions was randomized in each questionnaire phase.

Then, the photorealism of the environment was changed in a counterbalanced, randomized order and the embodiment phase and subsequent questionnaire phase were repeated three times so that each participant could evaluate the experience with all three rendering types of environment. The in-VR experiment took 15--25 minutes.

\subsubsection{Post-experiment Survey}
The participants revisited the website and filled out the post-experiment survey that consisted of demographic questions and an open-ended form about the experience.
After the completion of the protocol, the true purpose and hypotheses of the experiment were debriefed by e-mail if they wished.
In addition, if they applied for it, they were compensated with an Amazon gift card worth approximately \$10 after the completion of the application. The survey took around 5 minutes.

\section{Results}
\subsection{Data Collection and Cleansing}
\label{subsec:results-data}
The application was available for approximately one month, and 128 participants completed the experiment. Because data were collected remotely without experimenter supervision, where data collection is uncontrolled and protocol noncompliance is more likely than in laboratory settings \cite{Steed2016-bn}, we implemented prespecified data quality checks.

We applied two exclusion criteria and removed 11 participants in total (see Appendix for full details): 
(i) \textit{response reliability}, based on too-short completion times for answering questionnaires and response-pattern checks (e.g., invariant answers despite control items that were supposed to be answered inversely), and 
(ii) \textit{protocol compliance}, based on insufficient in-view time of the virtual hands during the embodiment phases as indicated by the tracking state data. Short out-of-view periods can occur naturally to some extent (e.g., resting hands), but this includes cases where hands were not detected properly due to the user's experimental environment not being optimized for tracking, or where the user did not display their hands as instructed. We excluded two and nine participants for criteria (i) and (ii), respectively.

The final analysis set comprised 117 participants, with a slight imbalance across hand conditions: realistic (38), sketch (40), and toon (39). The subsequent analyses in 4.2 use this sample.

\subsection{Questionnaire}
\figfloat{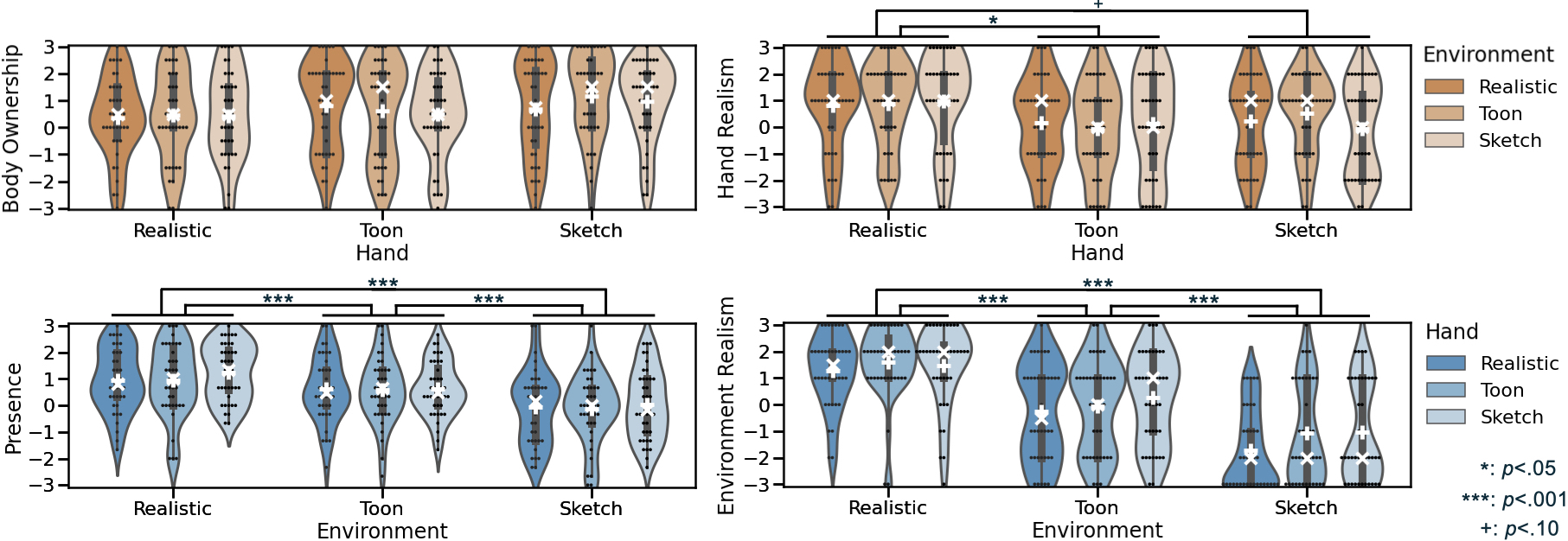}{Violin plots show the quartiles (vertical lines) and kernel density estimates of the distributions (outlines) of the questionnaire data. The black points represent all the observations. The $\times$ and $+$ marks refer to medians and means, respectively.
}
To test the hypotheses, the questionnaire data were analyzed.
For the scales that consisted of several items (sense of body ownership and presence), we first checked their internal consistency, after inverting the answers for negative items.
The Cronbach's $\alpha$ for body ownership was originally $\alpha=0.57$. As removing Q3 could increase $\alpha$ up to 0.70, we excluded Q3 from the body ownership score.
For presence, one item (Q6) was negatively correlated with the other presence items and was excluded from the calculation of the presence score.
Consequently, the Cronbach's $\alpha$ for presence rose to $\alpha=0.59$ from the original value ($\alpha=0.43$).
Then, for the sense of body ownership (Q1 and Q2) and presence (Q4, Q5, and Q7), the answers for each item were averaged
to compute the scores for each scale per participant.
Because a Likert scale is considered to be an ordinal scale, an aligned-rank transform, which allowed the use of ANOVA to analyze the interaction effects with non-parametric data~\cite{Wobbrock2011-fz}, was first applied, and then a two-way mixed ANOVA (between-subject factor of hand photorealism and within-subject factor of environment photorealism) was conducted for each scale (\autoref{fig:questionnaire_results}).
Below, we describe only the factors for which a significant difference was observed. See Appendix Table 2 for all the statistics.

For \textbf{the sense of body ownership}, no significant effects were observed.
For \textbf{presence}, the ANOVA revealed the significant main effect of the environment (\anovaETAbody{2}{228}{48.08}{<}{.001}{.297}), indicating a large effect size. 
Post-hoc pairwise comparisons
(Holm-corrected) showed that the scores in the realistic environment condition were significantly higher than those in the toon and sketch environment conditions and that the scores in the toon environment condition were significantly higher than those in the sketch environment condition (all {\it p}$<$.001).
For \textbf{hand realism}, the ANOVA revealed the significant main effect of the hand (\anovaETAbody{2}{114}{3.91}{<}{.05}{.064}), indicating a medium effect size.
Post-hoc pairwise comparisons
(Holm-corrected) showed that the scores in the realistic hand condition were significantly higher than those in the toon hand condition ({\it p}$<$.05).
The scores in the sketch hand condition did not significantly differ from those in the realistic and toon hand conditions ($p=.07$ and $p=.63$, respectively).
For \textbf{environment realism}, the ANOVA revealed the significant main effect of the environment (\anovaETAbody{2}{228}{131.04}{<}{.001}{.535}), indicating a large effect size.
Post-hoc pairwise comparisons
(Holm-corrected) showed that the scores in the realistic environment condition were significantly higher than those in the toon and sketch environment conditions and that the scores in the toon environment condition were significantly higher than those in the sketch environment condition (all {\it p}$<$.001).
Taken together, no significant interaction effects were observed for any of the scales.

\subsection{Post-experiment Survey}
The demographic distributions are summarized in \autoref{subsec:participants}.
Among the 122 participants who answered the post-experiment survey, 100 filled in the open-ended form about their experience.
Most participants (realistic: 19/40, toon: 17/42, sketch: 22/42) pointed out the low quality of hand tracking and subsequent failure in the pointing task. Some mentioned that it decreased the sense of immersion or presence to some extent when the tracking was lost. Others compared the influence of tracking quality with that of visual realism. For example, some representative quotes were as follows: \textit{"More than the quality of the visual experience ... the latency in the hand tracking ... made me more uncomfortable"} and \textit{"More than the realism of the hand and room, ..., it bothered me that the hand was not recognized when I placed it far away from me."}

Some participants (realistic: 11/40, toon: 3/42, sketch: 2/42) mentioned whether or not they felt like the virtual hands were their own hands.
Except for one positive comment \textit{"The palm lines were so close to the actual one that I wondered if it was generic or my own"} on the hand condition, all the comments were negative as follows: \textit{"The hands were too real that they [impaired the ownership/felt weird]"} (realistic), \textit{"It felt weird, like my hands weren't my own"} (realistic), \textit{"I don't like the hand because its color is too realistic and it looks like a cut hand"} (realistic), \textit{"The fingers were so small that I didn't feel that it was my hand"} (realistic), \textit{"I didn't think it was my hand because of [the delay of hand movements/tremor of the hands]"} (realistic), \textit{"The blue hand used in the questionnaire felt more like my own hand than the realistic skin-colored hand during the task"} (realistic), \textit{"I felt uncomfortable about [the shade of the hand/the black edges of the nails]"} (toon), and \textit{"I thought it would have been more realistic if the virtual hands were rendered in skin tones"} (sketch).

In addition, we added to the survey in the second half of the experimental period the following question: ``Which of the three scenes did you like best as an experience?''
Of the 42 valid answers, 30 (71.4\%) preferred the realistic environment, 11 (26.2\%) the toon environment, and one (2.4\%) the sketch environment.
In the open-ended question on the reason for their choice, the most frequent (13/30) description of the realistic environment was \textit{``It was close to reality''}, and the second most (8/30) was \textit{``It was realistic''} or \textit{``I felt reality.''} Two participants mentioned that \textit{``the sense of immersion was the highest.''}
Those participants who favored the toon environment mentioned task-related reasons (e.g., easy to find an object) most frequently (4/11).
The second most frequent answer (3/11) was that they liked the appearance (e.g., moderately virtual and stylish).
One participant who favored the toon environment stated that s/he chose this preference \textit{"because the depiction of the hands and the room matched."}
The participant who favored sketch commented that \textit{"the lack of color information made it look neat."}

\section{Discussion}
The main findings are as follows:
\textbf{1)} Neither the rendering style of the virtual hands nor the consistency of rendering style between the hands and environment (i.e., the interaction effect) affected the sense of body ownership over the virtual hands, which did not support H1 and H3.
\textbf{2)} The consistency of the rendering styles did not affect the sense of presence, which did not support H2.
\textbf{3)} The more realistically the virtual environment was rendered, the stronger the presence, which supported H4.

\subsection{General Discussion}
\textbf{Interpretation of the Results.} 
Contrary to our expectations, the sense of body ownership was overall similarly high in any combination of rendering styles.
Considering that photorealism is an element of realism, it is somewhat surprising that no independent effect of the rendering style of the virtual hand was even observed.
Indeed, the visual realism of the realistic virtual hand was evaluated higher than the other virtual hand types; hence, the manipulation of perceived realism was a success, at least in the realistic condition compared with the toon condition.
Therefore, our results indicate that the sense of body ownership did not necessarily increase even when the virtual hand was perceived as visually realistic.
This result is consistent with the study by Lin and J\"org~\cite{Lin2016-sb} in which participants' evaluation of realism for the realistic, toony, and very toony hands significantly differed, yet those of the sense of body ownership were similar. 

We consider three complementary explanations for the null effect of photorealism on body ownership, which yielded relatively high ownership scores (both mean and median above zero) across all conditions.
First, visuomotor synchrony, the primary bottom-up driver of the VHI, may have been sufficiently strong to elicit the illusion regardless of rendering style. Such bottom-up multisensory cues can dominate top-down modulation when the sensory evidence is compelling~\cite{Kilteni2012-ji}. In our experiment, the Quest's built-in hand tracking provided continuous visuomotor correspondence, which may have resulted in a ceiling-like effect across conditions. Note that the Leap Motion controllers were commonly employed in prior VHI studies to track hand movements (e.g., \cite{Schwind2017-hi,Lin2016-sb,Argelaguet2016-si}).
Second, photorealism may indeed exert only a weaker top-down influence on body ownership than anthropomorphism: in the study by Lin and Jörg~\cite{Lin2016-sb}, ownership was significantly lower only for the wooden block model, which lacked human-like morphology entirely, whereas virtual hands differing only in rendering style did not differ in ownership; taken together with our results, this suggests that anthropomorphic plausibility may be the more critical boundary for ownership to emerge among the dimensions of realism.

Third, the uncanny valley effect~\cite{McDonnell2012-oo,Schindler2017-sm,MacDorman2009-hz} may have counteracted any potential benefit of higher photorealism for the body ownership illusion in the realistic condition.
As reported in the post-experiment survey (Section~4.3), participants who used the realistic virtual hand frequently mentioned their feelings about the virtual hands in free-text responses: 11 out of 40 participants compared with 3/42 in toon and 2/42 in sketch. Many of these comments concerned discomfort with the hand's appearance, indicating that highly photorealistic rendering drew attention to fine-grained discrepancies between the virtual and real hand and possibly amplified perceptual mismatch. The uncanny valley was originally hypothesized regarding affective responses toward robots~\cite{Mori1970-vq}, but recent studies have pointed out that it may also manifest in body ownership illusion: non-human full-body self-avatars have been reported to elicit slightly stronger ownership than realistic human self-avatars along the anthropomorphism dimension~\cite{Lugrin2015-wx,Ogawa2020-da}. However, the relationship between uncanny valley effects and body ownership is not straightforward. Latoschik et al.~\cite{Latoschik2017-tv} reported some indication of uncanny valley effects for realistic photogrammetry-based avatars, yet those avatars evoked significantly higher body ownership than abstract wooden mannequin avatars, suggesting that negative affective reactions do not necessarily reduce ownership. Our data point in a similar direction: the realistic hand was felt realistic and elicited the most negative comments, yet did not produce lower ownership scores. Whether a non-monotonic pattern similar to the uncanny valley exists along the photorealism dimension remains an open question, and any extension of uncanny valley theory 
to the photorealism dimension requires careful empirical verification. Future work should directly measure both affective responses (e.g., eeriness) and ownership simultaneously to disentangle these constructs.

As for the influence of photorealism on presence, we found a large effect size.
Although the effect was controversial in earlier studies~\cite{Zimmons2003-cj,Mania2004-mm,Vinayagamoorthy2004-lk,Khanna2006-nu,Slater2009-ix, Yu2012-kl}, and even though a recent review summarized that the level of visual realism is not as important as might be imagined~\cite{Slater2020-xh}, the results of our relatively large controlled experiment support that rendering style affects presence, in line with several recent studies \cite{Schwind2019-th, Zibrek2019-xm,Zibrek2019-ha,Hvass2017-wy}.
As no interaction effects were found, our study primarily provides additional evidence on the main effects of photorealism on body ownership and presence. Nonetheless, given that prior evidence for these effects was inconclusive, and that VR technology has advanced considerably since those earlier studies, confirming them with contemporary off-the-shelf hardware and a relatively large sample constitutes a contribution, including reporting null effects, which are often underreported in this field.

\textbf{Design Implications.}
The results provide insights into the design of VR experiences from the perspective of rendering style.

\textit{General principle: independent optimization.}
Because no significant interaction effects of rendering style on either body ownership or presence were found, designers and developers can decide the rendering style of the body and environment independently.

\textit{Environments: prioritize photorealism when presence matters.}
Photorealism should be prioritized for the environment rendering, as the realistic environment elicited the highest sense of presence and was preferred by the most participants. This is particularly relevant for applications where presence is critical, such as professional training simulations (e.g., medical or industrial), therapeutic VR (e.g., exposure therapy), and virtual tourism, where the sense of ``being there'' is central to effectiveness.

\textit{Hands and bodies: photorealism is not a priority.}
There is no empirical evidence for prioritizing photorealistic rendering of virtual hands, considering that the realistic virtual hand tended to produce negative affective reactions according to qualitative data, while it did not significantly increase the sense of body ownership. This finding is particularly relevant for standalone VR headsets, where computational resources are constrained. In such cases, developers can allocate resources to environmental photorealism rather than hand photorealism without sacrificing body ownership. For social VR and collaborative applications, where real-time synchronization can be a priority under limited computational resources, stylized (e.g., toon-shaded) avatars may be a pragmatic choice that avoids negative reactions while preserving ownership.
Nevertheless, existing studies have shown that personalized full-body self-avatars increase body ownership more than generic photorealistic self-avatars~\cite{Waltemate2018-cf,Gorisse2019-zk}. For applications where strong body ownership is desired (e.g., embodied perspective-taking, rehabilitation), investing in personalization (truthfulness) may be a more effective strategy.

\textbf{Realism Classification.} 
We incorporated Garau~\cite{Garau2003-tf}'s definition of the visual fidelity of avatars into the classification of the realism of a virtual body in the BOI.
Distinguishing elements, rather than lumping them together as realism, may help resolve the apparent inconsistency in the evidence to date, and the impact of the appearance of virtual bodies on ownership could then be addressed more systematically.
Indeed, although the anthropomorphism of virtual hands has repeatedly been shown to influence the sense of body ownership (e.g., \cite{Argelaguet2016-si,Schwind2018-uu, Lin2016-sb,Yuan2010-ep,Ogawa2019-bw }), the impact of photorealism seems to be negligible; therefore, it is important to consider them separately.

Nevertheless, a classification for the top-down influence of BOIs already exists.
Kilteni et al.~\cite{Kilteni2015-jd} categorized the semantic constraints of a seen object in BOIs into the shape, texture, and anatomical plausibility of the spatial configuration and internal structure in their review article that covers various types of BOIs.
However, the photorealism dimension does not apply to any of these.
Texture, which would be the closest element, has been shown not to be crucial for BOIs but modulates the strength of the illusion; yet, the effect becomes negligible in the BOI for virtual bodies~\cite{Kilteni2015-jd, Peck2013-ea}.
Although this finding is consistent with our results when we consider texture as photorealism, we emphasize that photorealism and texture are not necessarily the same concept.
For instance, both photorealistic non-corporeal textures and non-photorealistic human-like textures can exist.
One study that focused on skin color showed that a full-body self-avatar with purple skin induced the BOI without reducing the strength of the illusion compared with a self-avatar with a realistic skin texture~\cite{Peck2013-ea}. Yet, there could be different rendering styles (i.e., photorealism) for the same skin color (i.e., texture).
Therefore, photorealism is an essential parameter for characterizing a virtual body despite being overlooked in conventional BOI research.

Moreover, Kilteni's~\cite{Kilteni2015-jd} classification does not include a dimension of truthfulness, perhaps because creating a personalized rubber hand in the rubber hand illusion is impractical.
Recently, Gorisse et al.~\cite{Gorisse2019-zk}, exploiting Garau's~\cite{Garau2003-ni} classification, studied the effect of the truthfulness of a third-person perspective full-body self-avatar on the sense of embodiment.
Interestingly, Schwind et al.~\cite{Schwind2017-hi} found that three levels of the deviation of virtual hands from real hands affect the feeling of presence: deviations from common human appearance, the user's gender, and the user's body.
In addition, several other studies focus on the visibility (e.g., \cite{Lugrin2018-sz, Ogawa2020-da}) and sensitivity to pain (e.g., \cite{Lin2016-sb}) of a virtual body.
Given these overlapping but not identical views, it is important to provide a framework that integrates the existing multiple classifications to systematize the impact of virtual body appearance on the BOI.

\subsection{Limitations}
First, the participants comprised only owners of HMDs, and the gender ratio was highly skewed. As noted in prior work, a characteristic challenge of in the wild experiments is user selection bias \cite{Steed2016-bn}. The participant pool may exhibit latent bias, for example an overrepresentation of early adopters of consumer HMDs, and the observed effects may be biased as a result. Indeed, remote VR studies tend to include a higher proportion of men, for example 81.8\% \cite{Saffo2021-oc}, due to the gender distribution among HMD owners. The reproduction of such bias may disadvantage underrepresented groups \cite{Peck2021-qs}, possibly overlooking gender differences observed in VHIs \cite{Schwind2017-hi}. We do not consider this ideal, and we regard the participant imbalance as an important limitation of this study. It should therefore be noted that our results may more strongly reflect the characteristics of HMD owners, who are likely more receptive to state of the art technology and more experienced with VR than the general population, which is more diverse in gender and in sensitivity to technology. Future research should complement our findings by controlled laboratory studies that adopt a balanced sample and explicitly examine gender differences.

In addition, based on the post-experiment survey comments and the tracking state analysis in the data cleansing process, it seemed that the hand tracking system did not work ideally.
Although the main conclusion of this study would not be significantly influenced by the potential tracking issues because the allocation of experimental conditions was fully randomized and counterbalanced, it should be noted that the hand tracking issue could considerably affect the user experience, especially body ownership, which could be negatively affected under all conditions equally.
It should also be noted that the temporal resolution (i.e., approximately 40 fps) did not reach the hardware's refresh rates, which could also compromise the overall ownership and presence, although there was an inherent trade-off between photorealistic and smooth renderings in practice.
Unlike controlled lab experiments, remote experiments are conducted in a variety of real environments, meaning that the environmental conditions vary a lot whereas they represent everyday VR experiences for consumers. 
The tracking issues seem to still have occurred despite providing specific prior-to-experiment instructions to the participants to eliminate as many factors as possible that could reduce the accuracy of hand tracking, such as the number of obstacles in the room, the lighting environment, the size of the room, and the wearing of clothes that could hide the shape of the hands.
Thus, it is necessary to consider that the findings are based on the practical experience that could be achieved with VR devices for consumers at the time at home.

Furthermore, we focused on the VHI as a form of BOI, rather than full-body BOI.
Researchers have traditionally used the VHI for BOI partly due to technological limitations but also because the hand is considered a socially and technologically specific body limb.
Therefore, although VHI is only a form of BOI, it has been considered and used as a representative of the BOI in previous studies (e.g.,~\cite{Argelaguet2016-si, Schwind2018-uu, Lin2016-sb}).
In addition, virtual hand interaction was common in state-of-the-art commercial VR equipment at the time of this study. Thus, part of the contribution of our research was to examine the user experience in situations such as those commonly experienced by general users.
However, the extent to which the findings from the VHI can be generalized to the full-body BOI will need to be carefully examined. Indeed, the visual discontinuity between the hand and limb can decrease body ownership~\cite{Tieri2015-pj}. 
In terms of photorealism, in particular, there may be an interaction in which the discontinuities are less acceptable in photorealistic virtual hands.

In this study, we examined how photorealism affects user experience in virtual reality by operationalizing realism for the virtual hand along a photorealism axis while holding anthropomorphism fixed, in contrast to prior work that has often varied anthropomorphism under the label of realism. 
Nonetheless, we acknowledge the limitation that the three rendering styles we used do not comprehensively span the latent space of photorealism. In fact, subjective ratings of hand realism did not differ significantly between the sketch condition and the other two conditions, indicating limited perceptual separation among the styles. Notably, the same manipulation of rendering style affected perceived realism for the environment and for the hand differently, suggesting that the manipulation of the rendering styles may produce smaller perceptual changes in virtual hand realism than in the virtual environment.

Finally, our findings are based on subjective evaluations using Likert scale questionnaires. Although questionnaires are a widely used approach to assess body ownership and presence, they carry inherent response biases, so objective and physiological measures can offer further complementary findings in future studies. In addition, as described in Section~3.5, we used a fourth, neutral rendering style during the questionnaire phase to avoid photorealism dependent influences on the response experience, for example, responding on a floating interface in a photorealistic scene with photorealistic hands and with fingers passing through buttons, and to provide a consistent response context across conditions. Nevertheless, introducing another style may have created an unintended confound and could compromise internal validity.

\subsection{Future Work}
Future research should include examining other elements of photorealism than rendering style (e.g., texture and shader), such as the mesh. Volkmann et al.~\cite{Volkmann2020-bg} investigated how the polygon counts of virtual objects, including the user's virtual hands, influence the sense of presence, concluding that a high polygon count might not be that crucial for presence, including self-presence. 
Whether this result is also applicable to the sense of body ownership in the VHI and whether there is an interaction between the polygon and rendering style needs to be examined.
It would also be interesting to explore the influence of rendering style in terms other than photorealism. We applied a sepia filter in our unofficial preliminary study. Since such image effects are usually applied to the entire image, a mismatch between the environment and the body's rendering style may easily affect the perception.
In addition, although environmental information did not affect the sense of body ownership in terms of the consistency of the rendering style, future research should explore the more cognitive influence of the environment (e.g., story and situation).

\section{Conclusion}
By synthesizing existing classifications of realism and applying this framework to the VHI context, we systematically examine the effects of photorealism as a dimension of realism on the VHI.
Specifically, we investigated whether the match between the rendering style of virtual hands and environments affects the sense of body ownership and presence, hypothesizing that environmental information could affect the BOI.
To this aim, we conducted a 3 $\times$ 3 mixed-design remote VR experiment (N=117) that factored in the rendering styles (i.e., realistic, toon, and sketch) of both the virtual hand and the environment.
The results suggested that neither the photorealism of virtual hands nor the consistency of photorealism influenced the sense of body ownership. Nevertheless, the more photorealistically the virtual environment is rendered, the stronger the presence.
Our relatively large controlled experiment provides additional insights into the independent effects of body and environmental photorealism on the sense of body ownership and presence, respectively, for which the evidence has thus far been inconsistent.
These findings have implications for VR design: developers can optimize the rendering style of virtual hands and environments independently, prioritize environmental photorealism for applications where presence is critical (e.g., training simulations and therapeutic VR), and adopt stylized hand rendering without compromising body ownership, a practical advantage for resource-constrained platforms such as standalone headsets.

%% file: 02_questionnaire_table.tex
\begin{table}[bht]
\caption{Questionnaire items used in the experiment. Items in italics represent the control questions.
Each response was scored on a seven-point Likert scale ($-3=$ not at all, $+3=$ very much).
}
\label{table:questionnaire}
\scalebox{0.9}[1]{
\begin{tabular}{|p{1.6cm}|p{0.43cm}|p{6.3cm}|}
\hline
Scale & Item  &  \\ \hline
Body\quad Ownership
& Q1 & Did you feel as if the virtual hands were your hands, with +3 being exactly like reality?\\        
& Q2& \textit{Did you feel as if the virtual hands were someone else?}\\ 
          & Q3& \textit{Did you feel as if you might have more than one body?}\\ \hline
Presence& Q4&Please rate your sense of being in the room you saw, where +3 represents your normal experience of being in a place.          \\
&Q5\protect\footnotemark[1]& To what extent were there times during the experience when the virtual reality became the 'reality' for you, and you almost forgot about the 'real world' in which the whole experience was really taking place? \\
&Q6\protect\footnotemark[2]&\textit{During the time of the experience, which was the strongest on the whole, your sense of being in the virtual room or of being in the real world?}\\
&Q7\protect\footnotemark[3]&\textit{During the time of the experience, did the virtual reality overwhelm you or did you often think to yourself that you were actually just sitting in your actual room wearing a headset?}\\
\hline
Hand\quad Realism& Q8 & How realistic did the virtual hands look, with +3 being exactly like reality?\\\hline 
Environment Realism  &  Q9& How realistic did the virtual room look, with +3 being exactly like reality?\\
\hline
\multicolumn{3}{l}{\protect\footnotemark[1]:$-3=$ not all, $+3=$ all the time}  \\
\multicolumn{3}{l}{\protect\footnotemark[2]:$-3=$ the virtual room, $+3=$ the real world}\\
\multicolumn{3}{l}{\protect\footnotemark[3]:$-3=$ VR overwhelmed, $+3=$ in your actual room}
\end{tabular}
}
\end{table}